\documentclass[12pt,a4paper]{article}
\usepackage{amsmath}
\usepackage{fullpage}
\usepackage{amsfonts}
\usepackage{graphicx}
\usepackage{epsfig}

\usepackage[latin1]{inputenc}

\begin{document}

\title{Comment on ``Large negative lateral shifts from the
  Kretschmann-Raether configuration with left-handed materials''
  Appl. Phys. Lett. 87, 221102, 2005}
\author{Antoine Moreau\\
LASMEA, UMR CNRS 6602, Universit\'e Blaise Pascal,\\ 24 avenue
  des Landais, 63177 Aubi\`ere, France.
\and
Didier Felbacq\\
GES UMR CNRS 5650, Universit\'e de Montpellier II, Bat. 21,
  CC074,\\ Place E. Bataillon, 34095 Montpellier Cedex 05, France.
}

\date{}

\maketitle

\abstract{The negative shifts seen by Wang and Zhu are not due to the
  excitation of surface plasmons but to leaky modes 
of the slab propagating backward. Provided the characteristics of the 
LHM slab are chosen correctly, it is shown that a leaky surface 
plasmon can actually be excited using the KR configuration.}\\

In a recent letter\cite{wang}, Wang and Zhu studied the negative
lateral shift of a beam reflected by a thin slab of left-handed 
material (LMH) in the Kretschmann-Raether (KR) configuration. 
They claim that an ``unusual standing wave'' is responsible
for this negative shift, and that this standing wave ``becomes
a surface wave''. However, the authors studied a structure in conditions that could in no
way allow to excite surface plasmons. For an interface between air and 
a LHM characterized by $\epsilon_2<0$ and
$\mu_2<0$, a surface plasmon can exist\cite{ruppin,shadrivov1} provided (i) the propagation
constant $\alpha$ along the interface is greater
than $\sqrt{\epsilon_2\,\mu_2}\,k_0$ and $k_0$, which means that the field is
evanescent in the two media and (ii) either $|\mu_2|<1$ and
$\epsilon_2\,\mu_2 >1$ (this case corresponds to a backward surface wave) or $|\mu_2|>1$ and $\epsilon_2\,\mu_2
<1$ (forward surface wave). The dispersion relation  of the surface
plasmon is given by 
\begin{equation}
\alpha_p = k \,\sqrt{\mu_2 \,\frac{\epsilon_2-\mu_2}{1-\mu_2^2}}.
\end{equation}

In their study of the KR configuration, Wang and Zhu (i) have
propagative waves in the LHM medium since they take $k_0<\alpha <
\sqrt{\epsilon_2\,\mu_2}\,k_0$ and (ii) have chosen $\epsilon_2=-5$ and
$\mu_2=-1$. It is completely impossible to excite a surface wave 
in these conditions.

\begin{figure}[h]
\centerline{\includegraphics[width=12cm]{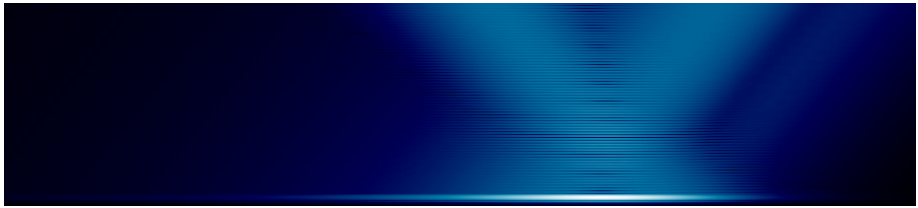}}
\caption{Modulus of the electric field when a leaky slab resonance is
  excited by an incident gaussian beam coming from the left.\label{f:slab}}
\end{figure}

We attribute the observed lateral shifts to leaky modes of the slab. It
has been shown that when a Fabry-Pérot resonance of the slab is excited
by an incident beam with an incidence angle $\theta \neq 0$, it gives birth
to a leaky mode of the slab\cite{pillon} although this is not well known. Leaky modes are responsible
for large lateral shifts\cite{tamir,pillon,chuang}. In our
case, the LHM slab supports a {\it backward} leaky mode\cite{tamir}, which
explains why the shift is found to be negative. The leaky mode can of course not be seen using a plane wave
analysis\cite{tamir}. Figure \ref{f:slab} shows the result of our simulations : an
incident beam with an incidence angle of 37.8° is reflected by
a LHM slab. The leaky mode is clearly propagating towards the left 
and its amplitude is exponentially decreasing in this direction.
This result is both new and interesting, but no surface resonance is
involved in this phenomenon.

\begin{figure}[h]
\centerline{\includegraphics[width=8.3cm]{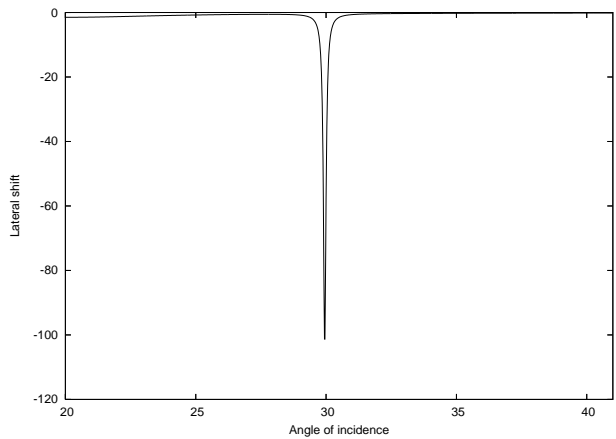}}
\caption{Lateral shift $\frac{\Delta}{\lambda}$\cite{wang,felbacq}
  versus the angle of incidence for $\epsilon_1=12$, $\mu_1=1$, $\epsilon_2=-5$ and
$\mu_2=-0.5$ and $d=0.6\,\lambda$. The angle above which the field is
  evanescent in the LHM is 27.16°. 
\label{f:shift}}
\end{figure}

Let us now consider a configuration for which a backward surface plasmon may
{\it a priori} be excited. We take $\epsilon_1=12$, $\mu_1=1$, $\epsilon_2=-5$ and
$\mu_2=-0.5$ and $d=0.6\,\lambda$. No slab resonance can be excited in
these conditions because the slab is too thin. The lateral shift for
this configuration is represented figure \ref{f:shift}. A very large
negative shift can be seen for $\theta\simeq 29.94$° when
$\alpha\simeq\alpha_p$. This shift is even greater than in the case of slab
resonances because the plasmon is less leaky and hence it propagates 
further. Let us stress that the field is evanescent in the LHM
($\sqrt{\epsilon_2\,\mu_2}\,k_0<\alpha$) and in
the air so that this resonance is without any doubt a surface plasmon.

We came to the conclusion that the negative shifts seen by Wang and
Zhu are not due to the excitation of surface plasmons but to leaky
modes of the slab\cite{tamir,pillon} propagating {\it backward} -
which is an interesting result. Provided
the characteristics of the LHM slab are chosen correctly, we showed it is 
possible to excite a leaky surface plasmon\cite{shadrivov2} using the KR 
configuration.

\newpage



\newpage

\begin{thebibliography}{20}

\bibitem{wang} L.G. Wang and S.Y. Zhu, Appl. Phys. Lett. {\bf 87}, 221102 (2005).

\bibitem{tamir} T. Tamir and H.L. Bertoni, J. Opt. Soc. Am. {\bf 61},
  1397 (1971).

\bibitem{pillon} F. Pillon, H. Gilles, S. Girard, M. Laroche, R. Kaiser and
A. Gazibegovic, J. Opt. Soc. Am. B {\bf 22}, 1290 (2005).

\bibitem{chuang} S.L. Chuang, J. Opt. Soc. Am. A {\bf 3}, 593 (1986).

\bibitem{ruppin} R. Ruppin, Phys. Lett. A {\bf 277}, 61 (2000).

\bibitem{shadrivov1} I.V. Shadrivov, A. Sukhorukov, Y. Kishvar,
  A. Zharov, A. Boardman, P. Egan,  Phys. Rev. E {\bf 69}, 016617
  (2004).

\bibitem{felbacq} D. Felbacq, A. Moreau and R. Smaali, Opt. Lett. {\bf 28}, 1633 (2003).

\bibitem{shadrivov2} I. Shadrivov, A. Zharov and Y. Kivshar,
  Appl. Phys. Lett. {\bf 83}, 2713 (2003). 

\end{thebibliography}
\end{document}